\documentclass[12pt]{article}
\usepackage{amsmath,amsfonts,bm}
\usepackage{authblk}

\usepackage{paralist}
\usepackage{comment}

\usepackage{caption}
\usepackage{array}

\usepackage{tikz}
\usetikzlibrary{arrows,matrix,backgrounds,shapes,positioning,fit,chains}

\newcommand{\code}[1]{\texttt{\small{#1}}}
\newcommand{\pkg}[1]{\textsf{#1}}
\newcommand{\bvec}{\left[\begin{array}{c}}
\newcommand{\evec}{\end{array}\right]}
\newcommand{\bmat}[1]{\left[\begin{array}{*{#1}{c}}}
\newcommand{\emat}{\end{array}\right]}

\newcommand{\dimm}[2]{\underset{\phantom{.}^{#2}}{#1}}
\newcommand{\inds}[2]{_{\genfrac{}{}{0pt}{}{#1}{#2}} }

\newcommand{\tr}{^\top}
\newcommand{\eps}{\varepsilon}
\newcommand{\R}{\textsf{R}}

\newcommand{\iid} {\operatorname{i.i.d.}}

\newcommand{\Var}{\operatorname{Var}}
\newcommand{\EV}{\operatorname{E}}
\newcommand{\Cov}{\operatorname{Cov}}

\newcommand{\diag} {\operatorname{diag}}

\newcommand{\snr}{\operatorname{SNR}}

\newcommand{\onevec}{\bm{1}}

 \newcommand{\mg}{\bm{g}}

 \newcommand{\mt}{\bm{t}}
 
 \newcommand{\mw}{\bm{w}}
 \newcommand{\mx}{\bm{x}}
 \newcommand{\my}{\bm{y}}
 \newcommand{\mz}{\bm{z}}
 \newcommand{\mA}{\bm{A}}
 \newcommand{\mB}{\bm{B}}

 \newcommand{\mE}{\bm{E}}

 \newcommand{\mI}{\bm{I}}
 
 \newcommand{\mK}{\bm{K}}

 \newcommand{\mP}{\bm{P}}

 \newcommand{\mS}{\bm{S}}

\newcommand{\mDelta}{\bm{\Delta}}
\newcommand{\mPhi}{\bm{\Phi}}
\newcommand{\mtheta}{\bm{\theta}}

\usepackage[bookmarks, colorlinks=TRUE, linkcolor=blue, urlcolor=blue,
citecolor=black, pdftitle={Functional Additive Mixed Models}, pdfauthor={Fabian Scheipl}]{hyperref}

\usepackage[paperwidth=8.5in, paperheight=11in]{geometry}
\usepackage[textsize=tiny, backgroundcolor=white,linecolor=red, bordercolor=red]{todonotes}
\usepackage{lineno}
\usepackage{setspace}

\definecolor{fgcolor}{rgb}{0,0,0}
\usepackage{framed}
\makeatletter
 {\par\unskip\endMakeFramed}
\makeatother

\definecolor{shadecolor}{rgb}{.97, .97, .97}

\usepackage{mathpazo}

\usepackage[sort&compress, authoryear]{natbib}
\begin{document}

\title{Functional Additive Mixed Models}
\author[1]{Fabian Scheipl}
\author[2]{Ana-Maria Staicu}
\author[1]{Sonja Greven}
\affil[1]{Ludwig-Maximilians-Universit{\"a}t M{\"u}nchen}
\affil[2]{North Carolina State University}

\maketitle

\begin{abstract}
We propose an extensive framework for additive
regression models for correlated functional responses, 
allowing for multiple partially nested or crossed functional random 
effects with flexible correlation structures for, e.g., spatial, temporal, or
longitudinal functional data. Additionally, our framework includes linear and
nonlinear effects of functional and scalar
covariates that may vary smoothly over  
the index of the functional response. It accommodates densely
or sparsely observed functional responses and predictors which may be observed
with additional error and includes both spline-based and
functional principal component-based terms.
Estimation and inference in this framework is based on standard additive mixed models, allowing us to take advantage of established methods and robust, flexible algorithms. 
We provide  easy-to-use open source software in the \code{pffr()} function for the \R-package
\pkg{refund}. 
Simulations show that the proposed method recovers
relevant effects reliably, handles small sample sizes well and also scales to larger data sets. 
Applications with  spatially and longitudinally observed functional data
demonstrate the flexibility in modeling and interpretability of results of our approach. 
\end{abstract}

\setstretch{1}

In recent years, many scientific studies have collected functional data that
exhibit correlation structures amenable to explicit modeling. 
Such structures may arise from a longitudinal study design
\citep[e.g.~][]{Morris:Carroll:2006, Greven2010, 
Goldsmith:etal:2012}, crossed designs \citep[e.g.~][]{Aston:etal:2009}, or
spatial sampling of curves \citep[e.g.~][]{Giraldo2010, Delicado2010, Nerini2010, Staicu2010, Gromenko2012}. 
Simultaneously, regression for independent functional responses
\citep[e.g.~][]{Faraway1997} has made large advances, 
including both multiple scalar \citep[e.g.~][]{Reiss2010}
and multiple functional predictors in concurrent or more general relationships
\citep[e.g.~][]{Ivanescu2011}. 
Our work is motivated by a longitudinal neuroimaging study containing repeated measurements of a 
functional proxy variable for neuronal health along 3 white matter tracts derived from 
diffusion tensor imaging (DTI). 
The goal of our analysis is to quantify the relationship of these function-valued  
proxy measures while accounting for the longitudinal correlation structure as well as the effects of patient characteristics like age and gender.
This DTI study is an example of a longitudinal functional data set where models
must account for the correlation structure of the data while including
both scalar and functional covariates in the predictor. 

To address these challenges, we propose conditional regression models for functional responses that 
accommodate general correlation structures via
functional and scalar random effects as well as flexible linear or nonlinear effects of
scalar and functional covariates.
The major contributions of this paper thus consist in 
1) developing a general inferential framework for additive mixed models for correlated functional responses that accommodates diverse correlation structures and flexible modeling of the mean structure extending \citet{Ivanescu2011},
2) 
unifying two previously separate strands of
prior work, by subsuming both functional principal component- (FPC) and spline-based approaches,
and 3) evaluating our implementation available in the
the \pkg{R}-package \pkg{refund} \citep{refund} on real and
simulated data.

Our goal is to describe and implement a framework that offers analysts
of functional data similar flexibility in model specification to what is
available in current implementations of (geo-)additive mixed models for scalar data.
Specifically, we consider structured additive regression models of the general
form
\begin{equation}
y_i(t) = \sum^R_{r=1} f_r(\mathcal{X}_{ri}, t) + \epsilon_{i}(t),
\label{eq:fdmodel}
\end{equation}
for functional responses $y_{i}(t), i=1,\dots,n,$ observed over a domain
$\mathcal{T}$.
Each term in the additive predictor 
is a
function of a) the index $t$ of the response and b) a subset $\mathcal{X}_r$ of the complete
covariate set $\mathcal{X}$ 
including scalar and functional covariates and
(partially) nested or crossed grouping factors. 
\begin{table}[htbp]
\begin{scriptsize}
\begin{center}
\begin{tabular}{p{.25\textwidth}|p{.35\textwidth}| p{.35\textwidth}}
$\mathcal{X}_r$ & $f_r(\mathcal{X}_{r}, t)$ constant over $t$ & $f_r(\mathcal{X}_{r}, t)$ varying over $t$ \\
\hline
&&\\
$\emptyset$ (no covariates)& scalar intercept $\alpha$ & functional intercept $\alpha(t)$\\[1em]
functional covariate $x(s)$ & linear functional effect $\int_{\mathcal{S}} x(s)
\beta(s) ds$ & linear functional effect $\int_{\mathcal{S}} x(s)
\beta(s, t) ds$ \\
& smooth functional effect $\int_{\mathcal{S}} F(x(s), s) ds$ & smooth functional effect $\int_{\mathcal{S}} F(x(s), s, t) ds$ \\[1em]
scalar covariate $z$ & linear effect $z\delta$ & functional linear effect $z\delta(t)$ \\ 
& smooth effect $\gamma(z)$  & smooth effect $\gamma(z, t)$ \\[1em]
vector of scalar covariates $\bm{z}$ & interaction effect $z_{1}z_{2}\delta$ & functional interaction effect $z_{1}z_{2}\delta(t)$ \\
					& varying coefficient $z_{1}\delta(z_{2})$ & functional varying coefficient $z_{1}\delta(z_{2}, t)$ \\  
				        & smooth effect $\gamma(\bm{z})$  & smooth effect $\gamma(\bm{z}, t)$ \\[1em]
grouping variable $g$ &  random intercept $b_{g}$ & functional random intercept $b_{g}(t)$ \\[1em]
grouping variable $g$ and 
scalar covariate $z$ &  random slope $z  b_{g}$ & functional random slope $z b_{g}(t)$
\end{tabular}
\end{center}
\end{scriptsize}
\caption{Forms of $f_r(\mathcal{X}_{r}, t)$ depending on the covariates in $\mathcal{X}_{r}$ and linearity or smoothness in these covariates (rows), and on whether the effect is constant  or varying over 
$t$ (columns). For scalar categorical covariates, synthetic scalar covariates in effect or reference category coding are created. 
Note that effects can become interaction effects if $\mathcal{X}_r$ additionally contains  
such scalar categorical covariates. For example, we estimate group-specific effects of the functional 
covariates for MS patients and healthy controls in our DTI application.}
\label{tab:effects}
\end{table}
To make this more concrete, Table \ref{tab:effects} shows the most important combinations
of $\mathcal{X}_r$ and effect shapes 
available in our framework. 
We assume a white noise error process independent of  $\mathcal{X}$, such that the $\epsilon_{i}(t)$ are independent and identically distributed (i.i.d.)~Gaussian variables with mean zero and
constant  variance $\sigma^2_\eps$ across $\mathcal{T}$.
Additionally, smooth and potentially correlated error curves can be included via curve-specific
random effects to model (co-)variance along $t$ and dependence between functional observations.
We assume all effects in Table \ref{tab:effects} to be smooth but unknown functions in the covariates, and this smoothness assumptions on all components of the predictor ensures smoothness of $y_{i}(t)$ up to the white noise measurement error $\epsilon_{i}(t)$. 
Scalar random effects $b_{g}$ are mean zero Gaussian
variables with general covariance structure between the different levels of $g$. 
Functional random effects $b_{g}(t)$ for a grouping variable $g$ with $M$
levels are modeled as realizations of a mean-zero Gaussian random process on
$\{1,\dots,M\} \times \mathcal{T}$ with a general covariance
function \mbox{$K^{b}(m,m',t,t') = \Cov(b_{m}(t), b_{m'}(t'))$} that is
smooth in $t$, where $m, m'$ denote different levels of $g$.
Note that our model class and software admits multiple partially or completely
nested or crossed grouping factors for both scalar and functional random
effects, but different random effects $b_{g}(t), 
b_{g'}(t)$ are assumed to be mutually independent. 
We assume integrability for the effects of functional covariates.  
Our implementation for functional effects such as $\int x_{i}(s) \beta(s,t) ds$
also accommodates varying integration ranges with fixed, potentially
observation-specific integration limits $l_i(t),u_i(t)$, similar to the historical functional model in \citet{MalfaitRamsay2003}. Densely as well as sparsely
observed functional responses and suitably preprocessed functional predictors with measurement error
can be used in this framework. We 
approximate each term $f_r(\mathcal{X}_r, t)$ by a linear combination of basis functions 
defined by the tensor product of marginal bases for $\mathcal{X}_r$ and $t$.
Since basis dimensions have to be sufficiently large to ensure enough flexibility,
 maximum likelihood estimation of model 
(\ref{eq:fdmodel}) is likely to lead to substantial overfitting.
The penalized likelihood approach described in Section \ref{sec:mixmodrep} stabilizes estimates by suppressing variability not strongly supported by the data and finds a data-driven
compromise between goodness of fit and simplicity of the fitted effects. 

Most existing work on functional random effects has considered only
special cases such as the functional random 
intercept model \citep{Abramovich:Angelini:2006,Di2009,Krafty:etal:2011},
functional random intercept and slope model \citep{Greven2010}, 
a single level of random effects functions  \citep{Guo:2002,Qin:Guo:2006,Antoniadis:Sapatinas:2007},
or a two or three-level hierarchy  \citep{Brumback:Rice:1998, Morris:etal:2003, Baladandayuthapani:etal:2008,
 Bigelow:Dunson:2007, Li:etal:2007, Scarpa:Dunson:2009, Zhou:etal:2010, Staicu2010}.
\citet{Aston:etal:2009}  consider a general functional random effects structure under the assumption of a joint functional principal component (FPC) basis for all functional random effects in the model,
which are estimated under a working independence assumption between curves. It is unclear, however, how well this approach works if the latent processes do not share the same eigenfunctions and how the correlation between functional observations affects FPC estimation. FPC estimation for
correlated observations is a topic of ongoing research \citep[c.f.][]{Hoermann:Kokozska:2010, Hoermann:Kokozska:2011, Panaretos:Tavakoli:2013a, Panaretos:Tavakoli:2013b}.
\citet{Morris:etal:2003,Morris:Carroll:2006,Zhu:etal:2011} propose a general
Bayesian functional linear mixed model based on a wavelet
transformation of (usually very spiky) data observed on an equidistant grid.  
The model proposed by \citet{Morris:Carroll:2006} includes correlation between different
random effects and heterogeneous residual errors, which we do not.  
Our approach, on the other hand, is well suited to smooth underlying curves and allows a more general mean structure than previous functional linear mixed models; in particular we are able to estimate smooth nonlinear or linear effects of scalar and/or functional covariates within the same framework. 
In addition, we are able to handle data on non-equidistant or sparse grids.

To the best of our knowledge, our proposal is the first publicly available
implementation that allows such a high level of flexibility for a functional regression
model -- prior work either limits the predictor to
the effect of a single functional covariate and a functional intercept, such as the \code{linmod} function in
package \pkg{fda} \citep{fda} for the \R~language \citep{R} or to linear
effects of scalar covariates, such as the \code{fosr} function for
function-on-scalar regression \citep{Reiss2010} in the \pkg{R}-package \pkg{refund}.
Like the linear function-on-function regression approach  in \citet{Ivanescu2011} we build on, both approaches are limited to independent functional responses.
\citet{Morris:Carroll:2006} provide a closed source implementation for wavelet-based
functional linear mixed models in \code{WFMM} \citep{WFMM} that allows very general random effect and residual structures, but implement neither effects of functional covariates nor nonlinear effects of scalar covariates. The \textsf{PACE} package \citep{PACE} for \textsf{MATLAB} implements FPC based regression 
models where the predictor is limited to the effect of a single functional or scalar covariate.
Our proposal has some similarities with the regression models  for independent or longitudinal scalar responses in \citet{Goldsmith:etal:2011,Goldsmith:etal:2012}, implemented in the 
\code{pfr} and \code{lpfr} functions in \pkg{refund},
since we also base  inference on additive mixed models for scalar-on-scalar regression. 
However, the extension to functional responses and functional random effects with 
flexible correlation structure as well as the inclusion of FPC-based
effects is non-trivial.

The paper is organized as follows:  Section \ref{sec:penreg} develops our general approach and estimation framework for functional additive mixed models. Our method is evaluated in a simulation study and in an application to the motivating
longitudinal DTI study 
in Section \ref{sec:empeval}. Section \ref{sec:finis} closes with a discussion and outlook. 


\section{Penalized regression for correlated functional data} \label{sec:penreg}


Functional responses $y_i(t)$ are observed on a grid
of $T_{i}$ points $\mt_{\bm{i}}=(t_{i1}, \dots,t_{iT_{i}})\tr$.
To simplify notation, 
we assume
identical grids $\mt_{\bm{i}}\equiv\mt=(t_{1},\dots ,t_{T})\tr$ for $i=1, \ldots, n$ in the
following, but note that functional responses observed on irregular and/or sparse grids
are naturally accommodated in the rephrased model formulation given in \eqref{eq:ammodel}. 
Then, model \eqref{eq:fdmodel} can be expressed as
\begin{equation}
\label{eq:ammodel}
  y_{il} = \sum^R_{r=1} f_r(\mathcal{X}_{ri}, t_l) + \epsilon_{il}
\end{equation}
for $i= 1, \dots, n$ and $l=1, \dots, T$.
The assumption of white noise errors translates to $\epsilon_{il}
\stackrel{\iid}{\sim} N(0, \sigma_\eps^2)$.
The smoothness assumption on $\EV(y_i(t))$ is preserved implicitly by
enforcing smoothness across $\mathcal{T}$ for all $f_r(\mathcal{X}_r, t)$.
To fit the model, we form 
$\my = (\my_{1}\tr, \dots, \my_{n}\tr)\tr$, an $n T$-vector that holds the concatenated function evaluation vectors
$\my_{i}=(y_{i1},\dots,y_{iT})\tr$. 
In the following, let $\bm{\mathcal{X}_r}$ denote the vector or matrix containing rows of observations $\mathcal{X}_{ri}$.
Let $f(\mt)$ denote the vector of function evaluations of  $f$ for each entry in the vector $\mt$ and let $f(\mx, \mt)$ denote the
vector of evaluations of   $f$ for each combination of rows in the vectors or matrices $\mx, \mt$. 
Let $\bm{1}_d=(1,\dots,1)\tr$ denote a $d$-vector of ones.
The row tensor product of an $m \times a$ matrix $\mA$ and an $m \times b$
matrix $\mB$ is defined as the $m\times ab$ matrix $\mA \odot \mB = (\mA \otimes
\bm{1}_b\tr) \cdot (\bm{1}_a\tr \otimes \mB )$, where $\cdot$ denotes element-wise multiplication. 

\subsection{Tensor product representation of effects}
\label{sec:penreg:tensorrep}
Each of the $R$ terms in model \eqref{eq:ammodel} can be represented as
a weighted sum of basis functions defined on the product space of the covariates in
$\mathcal{X}_r$ and $t$, where each marginal basis is associated with a corresponding
marginal penalty. A very versatile method to
construct basis functions on such a joint space is given by the row tensor product of marginal bases evaluated on $\bm{\mathcal{X}_r}$ and $\mt$
\citep[e.g.][ch. 4.1.8]{DeBoor78, Wood2006}.
Specifically, for each of the terms,
\begin{align}
\dimm{f_r(\boldsymbol{\mathcal{X}_r}, \mt)}{nT \times 1} &\approx \dimm{(\mPhi_{\bm{x}r}}{nT
\times K_{x}} \odot \dimm{\mPhi_{\bm{t}r})}{nT \times K_{t}} 
\dimm{\mtheta_r}{K_{x} K_{t} \times 1} = \mPhi_r \mtheta_r,
\label{eq:tensorrep}
\end{align}
$\mPhi_{\bm{x}r}$ contains
the evaluations of a suitable marginal basis for the covariate(s) in
$\boldsymbol{\mathcal{X}_r}$ and $\mPhi_{\bm{t}r}$ contains the evaluations of a marginal basis in $\mt$. The
shape of the function is determined by the vector of  coefficients $\mtheta_r$. A corresponding
penalty term can be defined by the Kronecker sum of the marginal penalty matrices $\mP_{\bm{x}r}$ and $\mP_{\bm{t}r}$ associated with each basis \citep[ch.
4.1]{Wood2006}, i.e.
\begin{align}
\operatorname{pen}(\mtheta_r| \lambda_{tr}, \lambda_{xr}) &=
\mtheta_r^T (\lambda_{xr}  \dimm{\mP_{\bm{x}r}}{K_x \times K_x} \otimes 
\mI_{K_t}+ \lambda_{tr} \mI_{K_x} \otimes \dimm{\mP_{\bm{t}r}}{K_t \times K_t})
\mtheta_r = \mtheta_r^T \mP_{r}(\lambda_{tr},\lambda_{xr}) \mtheta_r.
\label{eq:tensorpen}
\end{align}
$\mP_{\bm{x}r}$ and $\mP_{\bm{t}r}$ are known and fixed positive (semi-)definite penalty
matrices and $\lambda_{tr}$ and $\lambda_{xr}$ are positive smoothing
parameters controlling the trade-off between goodness of fit and the smoothness of 
$f_r(\boldsymbol{\mathcal{X}_r}, \mt)$ in $\boldsymbol{\mathcal{X}_r}$ and $\mt$, respectively. 
This flexible construction is valid for any combination of bases associated with  
quadratic penalties. 
Alternative constructions of the joint
penalty such as a direct Kronecker product $\lambda_r \left(\mP_{\bm{x}r} \otimes
\mP_{\bm{t}r}\right)$ associated with a single smoothing parameter $\lambda_r$  are
possible, see \citet[][ch. 4.1.8]{Wood2006} for a discussion. 
Typically, $K_x$ and $K_t$ vary with $r$ as well, but we drop the additional index for simplicity.
In the following paragraphs, we will motivate and define $\mPhi_{\bm{x}r},
\mPhi_{\bm{t}r}, \mP_{\bm{t}r}$ and $\mP_{\bm{x}r}$ for the different types of
terms available in our implementation. Effects that are constant
over $t$ are associated with $\mPhi_{\bm{t}r}= \bm{1}_{nT}$ and $\mP_{\bm{t}r} = \bm{0}$, 
while  users are free to choose any suitable
marginal basis matrix $\mPhi_{\bm{t}r}$ and penalty $\mP_{\bm{t}r}$
for terms that vary over $t$.

\paragraph{Spline basis representation of effects of scalar covariates}

For scalar covariates, index-varying effects are very similar to 
varying coefficient terms in models for scalar responses,
c.f.~\citet{Ivanescu2011}. For the functional intercept $\alpha(t)$,  
$\mPhi_{\bm{x}r}= \bm{1}_{nT}$ and $\mP_{\bm{x}r} = \bm{0}$.
For effects like $z \delta$ and $z\delta(t)$ that are linear in a scalar
covariate $z$, the marginal basis for the covariate direction reduces to
$\mPhi_{\bm{x}r}=\mz \otimes \bm{1}_T$ where $\mz = (z_1, \dots, z_n)\tr$, with penalty $\mP_{\bm{x}r} = \bm{0}$. For nonlinear effects of
scalar covariates like $\gamma(z)$ or $\gamma(z,t)$, 
$\mPhi_{\bm{x}r}$ is a suitable marginal spline basis matrix over $z$ and $\mP_{\bm{x}r}$ is
the associated penalty. 

\paragraph{Spline basis representation of functional effects}
\label{sec:penreg:integral} 

For linear effects of functional covariates  $x(s)$, we model $\beta(s, t)$ using 
tensor product splines with basis functions $\Phi_{k_s}(s), k_s=1,\dots,K_x$,
over $\mathcal S$ and a spline basis defined over $\mathcal T$. 
We approximate the integral by numerical integration on the grid defined by the
observation points  $s_1, \dots, s_H$ in $\mathcal S$. The
effect in
\eqref{eq:ammodel}  then is
\begin{align*}
\int_{\mathcal{S}} x_{i}(s)\beta(s, t_l)ds &\approx  \sum^{H}_{h=1} w_{h}
x_{i}(s_h) \sum^{K_x}_{k_s=1}\sum^{K_t}_{k_t=1} \Phi_{k_s}(s_h) \Phi_{k_t}(t_l)
\theta_{r, k_s k_t}. 
\end{align*}
In the notation of \eqref{eq:tensorrep},  
\mbox{$\mPhi_{\bm{x}r} = [\bm{x} \diag(\mw) \mPhi_s] \otimes \bm{1}_T \approx [\int_{\mathcal{S}} x_{j}(s) \Phi_{k_s}(s)
ds]\inds{i=1,\dots,n}{k_s=1,\dots,K_x} \otimes \bm{1}_T$,}
where $\mw = (w_1, \dots, w_H)\tr$  contains the quadrature weights for a numerical integration
scheme, $\mx = [x_i(s_h)]\inds{i=1,\dots,n}{h=1,\dots,H}$, and $\mPhi_{s} =
[\Phi_{k_s}(s_h)]\inds{h=1,\dots,H}{k_s=1,\dots,K_x}$.
$\mP_{\bm{x}r}$ in the tensor product penalty \eqref{eq:tensorpen} is
the penalty associated with the $\Phi_{k_s}(s)$. 
We can extend this construction, which is equivalent to the one introduced in \citet{Ivanescu2011}, 
to cover terms like $\int^{u_i(t)}_{l_i(t)} x_{i}(s) \beta(s,t) ds$ \citep[e.g.][]{MalfaitRamsay2003}
with fixed, potentially observation-specific integration limits $l_i(t),u_i(t)
\in \mathcal{S}$.
This is achieved by defining suitable weight matrices $\mw_{i,l}$ with zero
entries for $s_h < l_i(t_l)$ and $s_h > u_i(t_l)$.  
Such effects will often be required for covariates 
and responses that are observed on the same time domain, where responses
cannot be influenced by future covariate values. 
In the limit, this also includes the  concurrent model with terms
$X(t)\beta(t)$.

Our framework also extends to  
non-linear function-on-function effects $\int_{\mathcal{S}} F(x_{i}(s), s, t)
ds$, which  generalize the
functional generalized additive model  \citep{McLean:2012}
from scalar to functional responses. They offer
similar flexibility to purely nonparametric approaches like  
\citet{ferraty2006, ferraty2011, fda.usc}, e.g. In our
framework, such terms can be represented as
\begin{align*}
\int_{\mathcal{S}} F(x_{i}(s), s, t_l) ds &\approx  
\sum^{H}_{h=1} w_{h} \sum^{K_x}_{k_s=1}\sum^{K_t}_{k_t=1} \Phi_{k_s}(x_i(s_h),
s_h) \Phi_{k_t}(t_l) \theta_{k_s k_t} 
\end{align*}
with
$\mPhi_{\bm{x}r} = [(\bm{w}\tr \otimes \bm{I_n}) \bm{\Phi_s}] \otimes \bm{1}_T$
and
$\bm{\Phi_s} = [\Phi_{k_s}(x_i(s_h) , s_h)]\inds{i=1,\dots,n, h=1,\dots,H}{k_s=1,\dots,K_x}$,
and
$\mP_{\bm{x}r}$ the penalty associated with  $\bm{\Phi_s}$.
Basis functions $\Phi_{k_s}(x(s),s)$ can be tensor product basis functions
derived from marginal bases for $x(s)$ and $s$ or true bivariate basis functions.

\paragraph{FPC basis representation of functional effects}
\label{sec:penreg:ffpc} 
Consider a functional covariate expanded in the Karhunen-Lo{\`e}ve expansion
$x_{i}(s) = \sum_{k} \psi_k(s)\xi_{ik}$ with 
$\int \psi_k(s) \psi_{k'}(s)ds = \delta_{kk'}$; $\EV(\xi_{ik})=0$; $\Var(\xi_{jk})= \zeta_k$.
Under the assumption that \mbox{$\int_{\mathcal{S}} \sum_{k>K_x}
\psi_k(s)\xi_{ik} \beta(s, t)ds \approx 0$} for some $K_x$, i.e., that all
smaller modes of variation of $x(s)$ only have a negligible effect on $y(t)$, we can write 
\begin{align*}
\int_{\mathcal{S}} x_{i}(s)\beta(s, t)ds &\approx
\int_{\mathcal{S}} \sum^{K_x}_{k=1} \psi_k(s)\xi_{ik} \beta(s, t)ds = \sum^{K_x}_{k=1} \xi_{ik} \tilde \beta_k(t) 
\end{align*}
with
$ \tilde \beta_k(t) = \int_{\mathcal{S}} \psi_k(s)\beta(s, t)ds$.
Thus, a linear function-on-function effect can be represented as a
sum of varying coefficient terms for the FPC loadings $\xi_{ik}$. 
This representation extends FPC regression approaches \citep[e.g.][]{ReissOgden2007}
from scalar  to functional responses. 
In the notation of the general framework,  
$\mPhi_{\bm{x}r} =
\left[\hat\xi_{ik} \right]\inds{i=1,\dots,n}{k=1,\dots,K_x} \otimes \bm{1}_T$ and
$\mP_{\bm{x}r} = \bm{0}$.  For the $t$-direction, $\mPhi_{\bm{t}r}$ and
the associated $\mP_{\bm{t}r}$ can be chosen freely. 
An implicit assumption here is that all $\tilde \beta_k(t), k=1,\dots,K_x$
have similar smoothness, as they are all associated with the same smoothing
parameter. 

This FPC-based approach may be advantageous for
functional covariates observed on irregular or sparse grids -- for such data,
the spline-based method requires a preprocessing step \citep[c.f.][]{James2002, Goldsmith:etal:2011, Goldsmith:etal:2012} to impute the incomplete trajectories on a dense and regular grid, whereas FPCs can be estimated directly from sparse data \citep{Yao2005}.  
Additionally, if the shapes of the $\hat\psi_k(s)$ are meaningful to
practitioners and $K_x$ is small, the coefficient functions $\tilde \beta_k(t)$ may be easier to interpret
than a coefficient surface $\beta(s, t)$. On the other hand,
since inference is performed \emph{conditional} on the estimated FPCs
$\hat\psi_k(s)$ and associated loadings $\hat\xi_{ik}$, coverage issues 
associated with these neglected sources of estimation variability 
\citep[c.f.][]{Goldsmith:Greven:Cranic:2012} and bias introduced by estimation
error in the FPC analysis step may occur. Additionally, 
$\int_{\mathcal{S}} \sum_{k>K_x} \psi_k(s)\xi_{ik} \beta(s, t)ds \approx 0$
might be a strong assumption that is hard to check in applications, as is the
choice of the discrete tuning parameter $K_x$. 
 
As in the spline-based case, we can extend this to FPC-based nonlinear
function-on-function effects.
The proposal corresponds  to an extension of the functional
additive model  by \citet{Muller2008} from scalar to functional responses,  with the effect of the functional
covariate  given by $f_r(x_i(s),t) = \sum^{K'_x}_{k=1} F_k(\hat\xi_{ik}, t)$.
In the notation of our general framework, 
$\mPhi_{\bm{x}r} = \left[\mPhi_{{\xi_1}}|\dots|\mPhi_{{\xi_{K'_x}}}\right]$, 
where $\mPhi_{{\xi_k}} =
\left[\Phi_a(\hat\xi_{ik})\right]\inds{i=1,\dots,n}{a=1,\dots,A} \otimes \bm{1}_T$ for suitable
spline basis functions $\Phi_a(\cdot)$, such that $K_x =  A K'_x$.
The marginal penalty is given by $\mP_{\bm{x}r} = \mI_{K'_x} \otimes \mP_\xi$,
where $\mP_\xi$ is the penalty associated with the $\Phi_a(\hat\xi_{jk})$.
In $t$-direction, we are again free to choose any suitable basis
$\mPhi_{\bm{t}r}$ and penalty $\mP_{\bm{t}r}$. Further extensions to interaction effects of estimated FPC scores 
$f_r(x_i(s),t) = \sum^{K'_x}_{k=1}\sum_{k < k' \leq K'_x} F_{k,k'}(\hat\xi_{ik},\hat\xi_{ik'}, t)$ 
are also obvious in this framework.\\

\paragraph{Spline basis representation of functional random effects} 
\label{sec:penreg:funrandom}

Functional random effects $b_{g}(t)$ are represented as
smooth functions in $t$ for each level $1, \dots, M$ of the grouping variable $g$. 
In the notation of equation \eqref{eq:tensorrep}, functional random intercepts
 are
associated with a marginal basis $\mPhi_{\bm{x}r} =[\delta_{g(i)
m}]\inds{i=1,\dots,n}{m=1,\dots,M} \otimes \onevec_T$, where $g(i)$ denotes the level of $g$ for observation $i$.
This yields an incidence matrix mapping the observations to the different levels of the grouping variable.
For a functional random slope effect in a scalar covariate $z$,
$\mPhi_{\bm{x}r}  = [z_{i} \delta_{g(i) m}]\inds{i=1,\dots,n}{m=1,\dots,M}\otimes \onevec_T$.
In the notation of equation \eqref{eq:tensorpen}, the marginal penalty $\mP_{\bm{x}r}$ for
functional random effects is a $M \times M$ precision matrix that defines the
dependence structure between the levels of $g$.
The quadratic penalty \eqref{eq:tensorpen} is mathematically equivalent to the 
distributional assumption $\mtheta_r \sim N(\bm 0,
\sigma^{-2}_\epsilon \mP_{r}(\lambda_{tr},\lambda_{xr})^{-1})$
\citep{Brumback:1999}. 
Through the representation in  \eqref{eq:tensorrep},  this induces a
mean zero Gaussian process assumption $b_{g}(t)\sim GP(0, K^{b}(g(i), g(i'),
t,t') )$, with covariance evaluated for all $nT$ observations $K^{b}\left(\mg, \mg, \mt,
\mt\right) = \sigma^{-2}_\epsilon \mPhi_r
\mP_{r}(\lambda_{tr},\lambda_{xr})^{-1} \mPhi_r\tr$.
The smoothing parameter $\lambda_{xr}$ 
controls the relative contribution of the inter-unit variability relative to 
the common roughness of the functional random effects controlled by
$\lambda_{tr}$.

If observations on different levels of the grouping factor are assumed
independent, $\mP_{\bm{x}r}=\mI_M$ is simply the identity matrix. 
More generally, $\mP_{\bm{x}r}$ can represent any fixed dependence
structure between levels of $g$:  It can be a (partially improper) precision matrix of a random
field with known correlation structure, implied, for example, by the spatial or
temporal arrangement of the different levels of $g$, such as a Gaussian Markov random field (GMRF) on
geographical regions for conditionally auto-regressive (CAR) model terms. 
Alternatively, $(\mP_{\bm{x}r})^{-1}$ can be defined using any valid correlation
function based on -- for example -- spatial,   
temporal, or genetic distances between levels of $g$. 
If the grouping variable is simply the index of observations (i.e., $g(i)=i$),
this construction yields smooth residual curves with potential for spatial or temporal autocorrelation. 
This innovative definition of functional random effects admits very flexible
model specifications, since any combination of spline basis, smoothness penalty and between-subject correlation can 
be used for functional random effects. This allows,
for example, for spatially correlated functional residuals with periodicity
constraints for the Canadian Weather data (see Appendix C of the online supplement). 
Multiple (partially) nested or crossed random effects can be constructed in this way and are implemented in \code{pffr()}. 

\paragraph{FPC basis representation of functional random intercepts}
\label{sec:penreg:pcre} 
For functional random intercepts without between-unit correlation, i.e., for
$b_{g}(t) \stackrel{\iid}{\sim}GP(0, K^b(t,t'))$, it can be advantageous to use 
the eigenfunctions of the covariance operator $K^b(t,t')$ as basis functions in $t$.
Specifically, we use the Karhunen-Lo{\`e}ve
expansion of random processes to represent $b_{g}(t) \approx \sum^{K_t}_{k=1}
\eta_k(t)\nu_{g k}$ with $\kappa_k, \eta_k(t)$ the ordered 
eigenvalues and -functions of $K^b(t,t')$,  $\nu_{g k}$ the
associated FPC loadings, and $K_t$ a suitable truncation lag.
The marginal basis for the $t$-direction is then 
$\mPhi_{\bm{t}r}= \bm{1}_n \otimes [\hat\eta_1(\mt)| \cdots|\hat\eta_{K_t}(\mt)]$.
Since $\EV(\nu_{{g} k})=0$ and $\Var(\nu_{{g} k})= \kappa_k$, a reasonable
marginal penalty is $\mP_{\bm{t}r}= \diag(\hat\kappa_1,
\dots,\hat\kappa_{K_t})^{-1}$. This encourages relative contributions of
the FPCs to the random effect curves that are roughly proportional to
their estimated magnitudes $\hat\kappa_k$. As for the spline-based functional random effects,
$\mPhi_{\bm{x}r}=[\delta_{g(i) m}]\inds{i=1,\dots,n}{m=1,\dots,M} \otimes \onevec_T$ is an
incidence matrix for the group levels, while $\mP_{\bm{x}r}=\mI_M$.

In practice, $\eta_k(t)$ and $\kappa_{k}$ have to be estimated. An iterative procedure
can be outlined as follows: 
(1) Use a fit without functional random effects under independence assumption
to obtain working residuals $\mE=[\eps_{il}]\inds{i=1,\dots,n}{l=1,\dots,T}$.
(2) Compute the group-level means of residual curves $\bar\mE= \mDelta\tr\mE$ 
with $\mDelta=[\delta_{g(i)m}/n_m]\inds{i=1,\dots,n}{m=1,\dots,M}$, where $n_m$
is the number of observations for the $m$-th level of $g$. (3) Perform a (truncated) spectral decomposition of $\hat \mK^b = [\hat
K^b(t_l, t_{l'})]_{l,l'=1,\dots,T}$ to obtain 
$\hat\eta_k(\mt),\hat\kappa_k$ for $k=1,\dots,K_t$. A suitable estimate 
for $\hat \mK^b$ can be derived from smoothing the entries in the matrix
$M^{-1}\bar\mE\tr\bar\mE$ (without the diagonal) as  in
\citet{Yao2005}.
This approach for estimating $\hat \mK^b$ can only be used for random intercepts
for a single grouping variable.
Compared to spline-based functional random effects, FPC-based modeling holds 
the promise of using the optimal, most parsimonious basis to represent $b_{g}(t)$. 
Computationally, it is expected
to scale much better for large  $M$, as the number of coefficients associated
with a functional random effect  is $M K_t$ and $K_t$ for FPCs will
typically be much smaller than in  a
sufficiently flexible spline basis. 
On the other hand, the FPC approach requires a pilot estimate for $\hat \mK^b$.
The subsequent performance  will be sensitive to the quality of the estimation of the  FPCs and to the choice of $K_t$.

\subsection{Mixed model representation}\label{sec:mixmodrep}

Using the tensor product representation introduced  in the previous subsection for 
 all terms,  model  \eqref{eq:fdmodel}  can be re-written as
\begin{align}
\my &=  \mPhi \mtheta + \bm{\epsilon}; \qquad
\bm{\epsilon} \sim N(\bm{0}, \sigma^2_\epsilon \bm{I}_{nT}),
\label{eq:lpffr_main_splines}
\end{align}
where $\mPhi = [\mPhi_1 | \dots | \mPhi_R]$ contains the  concatenated  
 $\mPhi_r$ associated with the different model terms
and $\mtheta=(\mtheta_1\tr, \dots ,\mtheta_R\tr)\tr$
 the respective stacked coefficient vectors $\mtheta_r$. 
To clear up notation, we assign a sequential index $v = 1, \dots, V$ to the smoothing parameters
$\lambda_{xr}, \lambda_{tr}$ in \eqref{eq:tensorpen}, where $V$ is the total number of smoothing parameters in the model.
 We pad $\mP_{\bm{t}r} \otimes 
\mI_{K_x}$ and $\mI_{K_t} \otimes \mP_{\bm{x}r}$ with rows and
columns  of zeros, denoting these matrices  $\tilde\mP_{v_1}$ and $\tilde\mP_{v_2}$,
such that the penalty $\mtheta_r^T (\lambda_{tr}  \mP_{\bm{t}r} \otimes 
\mI_{K_x} + \lambda_{xr} \mI_{K_t} \otimes \mP_{\bm{x}r})
\mtheta_r = \lambda_{v_1} \mtheta^T \tilde\mP_{v_1} \mtheta +  
\lambda_{v_2} \mtheta^T \tilde\mP_{v_2} \mtheta$ refers to the
full coefficient vector $\mtheta$. 
The penalized likelihood criterion to be
minimized then becomes
 \begin{eqnarray}
  \frac{1}{\sigma^{2}_\epsilon}\|{\my} -  \mPhi \mtheta\|^2 +
  \sum^{V}_{v=1} \frac{\lambda_v}{\sigma^{2}_\epsilon}\mtheta^T \tilde\mP_v
  \mtheta.
 \label{eq:penlikel}
 \end{eqnarray}
The total number of smoothing parameters is $V \leq 2R$, as some terms are constant over $t$ or $\mathcal{X}_r$  and the corresponding 
  $\tilde\mP_v$ are zero.
  Let $\tau_v= \tfrac{\sigma^{2}_\epsilon}{\lambda_v}$ and use
  similar arguments as in \citet[][ch. 4.9]{Ruppert2003} to obtain the solution
  $\widehat\mtheta$ of \eqref{eq:penlikel} as the best linear unbiased predictor in the linear mixed effects model (MEM)
 \begin{eqnarray}
 \my
 \sim N\left(\mPhi \mtheta, \sigma^2_\epsilon \mI_{nT} \right); \qquad
 {\mtheta} \sim N\left( \bm 0, \left(\sum^{V}_{v=1} \tau^{-1}_v \tilde\mP_v
 \right)^{-} \right),
 \label{eq:MEM}
 \end{eqnarray}
 where $\mS^{-}$ denotes the generalized inverse of $\mS$, and $N(\bm{0},
  \mS^{-} )$ is a partially improper Gaussian distribution with positive
  semi-definite covariance matrix $\mS$. The impropriety results from rank
  deficiencies in some of the $\tilde\mP_v$, since roughness penalties
  typically define a nullspace of maximally smooth functions. Numerical
  difficulties posed by the positive semi-definiteness are solved by 
  another re-parameterization that separates the various model terms
  into their unpenalized and penalized components, i.e. into ``fixed'' effects
  and ``random'' effects with a proper distribution, respectively.
  These are well known issues in the literature on
  penalized regression splines described in detail e.g.~in \citet[][ch. 4.9]{Ruppert2003},\citet[][ch. 6.6.1]{Wood2006}; recent developments for tensor product splines are in \citet{wood2012}.
  
One of the main advantages of formulating the penalized likelihood optimization
as estimation in an MEM is that the smoothing parameters $\lambda_v =
\tfrac{\sigma_\epsilon^2}{\tau_v}$ can be
treated as variance component parameters  and thus can be estimated using restricted
maximum likelihood (REML). In particular, \citet{Reiss:Ogden:2009} and
\citet{Wood2011} have shown that smoothing parameter selection with REML is
more stable and results in somewhat lower MSE than generalized cross-validation (GCV) and
\citet{Krivobokova:Kauermann:2007} have shown that REML estimation of penalized splines is more robust to error correlation
mis-specification than AIC-based criteria. A second advantage this approach offers is that the
representation of our model class \eqref{eq:fdmodel} results in a
fit criterion \eqref{eq:MEM} equivalent to that of conventional additive mixed models for scalar data. 
This means much of the powerful and versatile inference machinery developed for scalar linear and additive
mixed models (AMMs)  over the last years can be applied directly to the proposed model class of
functional AMMs, due to their close structural similarity. 
Specifically, 1) pointwise, bias-corrected
confidence bands \citep{Nychka1988, Ruppert2003, Marra2012} are available for
the functional effects, 2) tests for random effects as well as tests
for constant or linear effects versus more general alternatives developed for
scalar responses \citep{Crainiceanu:Ruppert:2004, Crainiceanu2005, greven2008,
scheipl2008, wood2012p}, 
and 
3) model selection approaches that have recently
been proposed for scalar-response AMMs \citep{Greven:Kneib:2010, Marra2011} are
directly applicable to the proposed model class. 
Finally, the proposed approach accommodates a large variety of
effects, at no increase in the level of complexity of the algorithm itself. 
The tensor product representation given in Section \ref{sec:penreg:tensorrep}
combined with the MEM representation \eqref{eq:MEM}
allows for a unified framework for smoothness parameter selection and estimation
of all model components in model \eqref{eq:fdmodel}, including functional
random effects and FPC-based effects. 

\paragraph{Implementation}

The full framework for functional additive models we describe here is
implemented in the \code{pffr}-function in the \pkg{refund} package for \R. 
The underlying inference engine is the \pkg{mgcv} package \citep{Wood2011} for
generalized additive models which also supplies most of the functionality for
constructing basis and penalty matrices.  
\code{pffr} offers a formula-based interface similar
to the established formula syntax of \pkg{mgcv} and returns a rich model object
whose fit can be summarized, plotted and compared with other model formulations without
any programming effort by the user through convenient utility functions.

\section{Empirical evaluation}\label{sec:empeval}
The following section describes an extensive simulation study and results for the motivating application to a longitudinal DTI study. A fully reproducible example analysis of the well 
known Canadian Weather data showcasing the flexibility of \code{pffr} can be found in Section C of the online appendix.
\subsection{Simulation study}\label{sec:empeval:sim}

\paragraph{Simulation setup}
We simulate data with repeated measures structure for a model \linebreak\mbox{$y_{ij}(t) =
\sum^R_{r=1} f_r(\mathcal{X}_{rij}, t) + \eps_{ij}(t)$} for the following four scenarios to investigate the sensitivity of the estimates to varying model complexity, noise levels and number of observations:
\begin{compactenum}[1.]
\item Functional random intercept, functional random slope: $\sum f_r(\mathcal{X}_{rij}, t) =
\alpha(t) + b_{i0}(t) + b_{i1}(t)u_{ij}$
\item Functional random intercept, functional covariate:\\ $\sum f_r(\mathcal{X}_{rij}, t)
= \alpha(t) +\int x_{1, ij}(s)\beta_1(s,t)ds + b_{i0}(t)$
\item Functional random intercept, two functional covariates:\\ $\sum f_r(\mathcal{X}_{rij}, t)
= \alpha(t) +\int x_{1,ij}(s)\beta_1(s,t)ds + \int x_{2,ij}(s)\beta_2(s,t)ds +
b_{i0}(t)$
\item Functional random intercept, functional covariate, smooth
scalar covariate effect, varying coefficient term: $\sum f_r(\mathcal{X}_{rij}, t))
= \alpha(t) +\int x_{1,ij}(s)\beta_1(s,t)ds + \gamma_1(z_{1, ij}, t) +
\delta_2(t) z_{2, ij} + b_{i0}(t)$
\end{compactenum}
Definitions of the various effect functions and descriptions of the data generating
processes used for the covariates can be found in section B of the online
supplement, along with unabridged simulation results and graphical displays of data 
and estimated effects for the replications with minimal, maximal and median error for each scenario.

\begin{figure}[!ht]
\begin{center}
  \includegraphics[width=\textwidth]{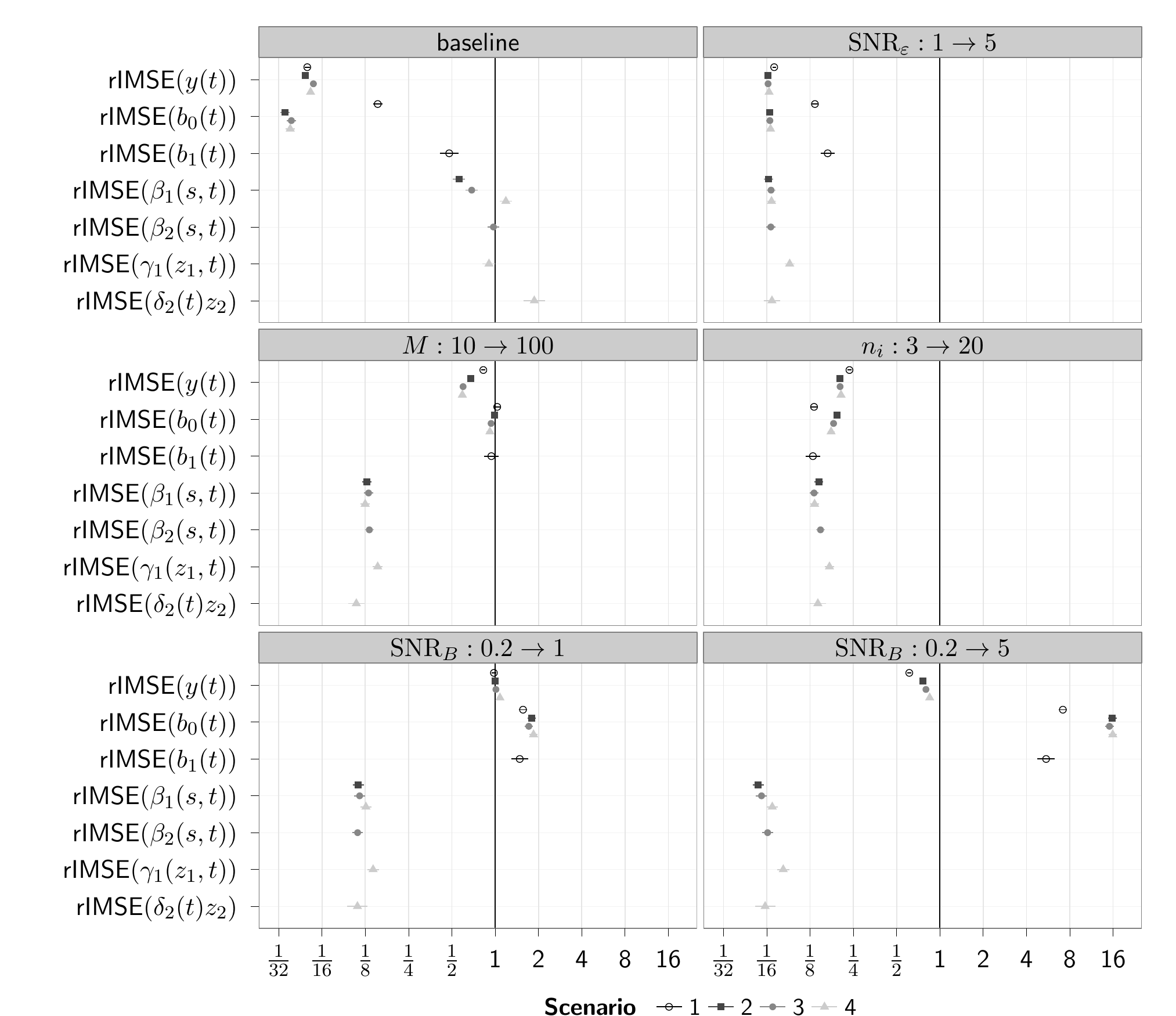}
  \caption{Baseline levels and estimated multiplicative change in rIMSE for
  the 4 scenarios. The scenarios are depicted with different symbols, 
  and the segments accompanying the symbols correspond to the estimated effect
  $\pm$ 2 standard errors. Effects other than $b_0(t)$ only occur in a subset of
  scenarios.
  Horizontal axis on $\log_{2}$-scale. }
  \label{fig:sim:coefplot}
\end{center}
\end{figure}

\begin{figure}[!hb]
\begin{center}
  \includegraphics[width=\textwidth]{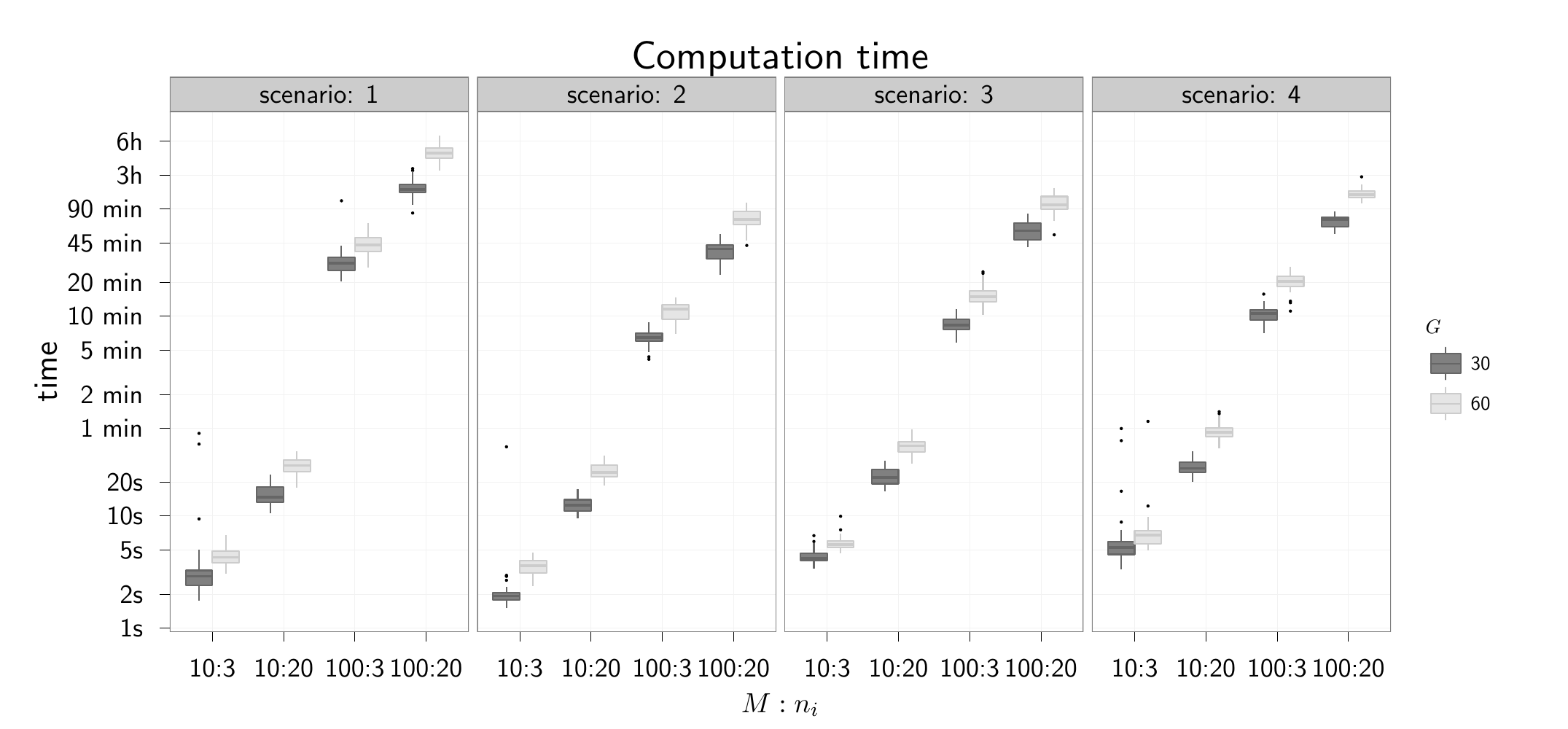}
  \caption{Computation times for scenarios 1 to 4 (from left to right).
  Vertical axis on $\log_{10}$-scale.
  Horizontal axis for the various combinations of numbers of subjects $M$ and average number of replications per subject $n_i$. Results for
  $T=30$ in dark grey and in light grey for $T=60$. Timings are wall-clock time taken on an 2.2 GHz
  AMD Opteron 6174.}
  \label{fig:sim:comptimes}
\end{center}
\end{figure}

For each of the four scenarios, we run 10 replications for each combination of
the following settings, yielding 1920 model fits in total:
\begin{compactitem}[--]
  \item number of subjects: $M \in \{10, 100\}$
  \item mean number of observations per subject: $n_i \in \{3, 20\}$. 
  Subject labels $i \in \{1,\dots,M\}$ are drawn from a multinomial distribution
  with probabilities $P(i=i') \propto \sqrt{i'}$ to generate unbalanced 
  designs. 
  \item number of grid points for $t$: $T\in \{30, 60\}$
  \item relative importance of random effects: $\snr_B\in \{0.2, 1, 5\}$, where $\snr_B$
  is the ratio of the standard deviation of the additive predictor without
  random effects divided by the standard deviation of the random effect
  functions; e.g.~for
  $\snr_B=5$, the contribution of each functional random effect to the variability in $y(t)$ is about 5 times smaller than that of the non-random effects.
  \item signal-to-noise ratio: $\snr_\eps\in \{1, 5\}$, where $\snr_\eps$
  is the ratio of the standard deviation of the additive predictor
  divided by the standard deviation of the residuals $\sigma_\eps$.
\end{compactitem}
Our results show that 10 replications for each combination are sufficient to
derive precise estimates of effects of the setting parameters on estimation
errors and computation times, c.f.~Figures \ref{fig:sim:coefplot} and
\ref{fig:sim:comptimes}. Fits are obtained with the defaults  in
\code{pffr()}, i.e., cubic B-spline bases with 20 basis functions and first
order difference penalty for the functional intercept, tensor products of cubic
B-spline bases with five marginal basis functions for the tensor product terms,
with first order difference penalties for the $t$- and $s$-directions
and second order difference penalties for the covariate direction (if
applicable). The smoothing parameters are REML estimates as returned by
\pkg{mgcv}.
The models for settings 1 to 4 include $K=120$ to 1020 
coefficients and 5 to 8 smoothing parameters. 

\paragraph{Estimation error}
We use the relative integrated mean squared error defined as \linebreak
rIMSE$(\hat f_r(\mathcal{X}_r, t)) = n^{-1} \sum^n_{i=1}\tfrac{\int (\hat
f_r(\mathcal{X}_{ri}, t) -f_r(\mathcal{X}_{ri}, t))^2 dt}{\int
f_r(\mathcal{X}_{ri}, t)^2dt}$ to evaluate the accuracy of the estimates.
Relative errors allow comparisons  across different scenarios and noise levels regardless of the
ranges of the true $f_r(\mathcal{X}_{r}, t)$. Note that we evaluate
the estimation accuracy of the effects on the scale of the response, not on the
scale of the coefficient function itself to make errors directly comparable
across effects. 
Detailed analysis of results (see Appendix B in the supplement)
shows that there are no relevant interaction effects between the setting
parameters $M$, $n_i$, $T$, $\snr_B$ and $\snr_\eps$ on the
observed errors within scenarios, so we fit log-linear models
with main effects for the setting parameters in each scenario to
observed rIMSE values and proceed to analyze the estimated effects. 
Figure \ref{fig:sim:coefplot} shows baseline levels and the estimated
multiplicative effects of the simulation parameters on the rIMSEs. 
The effect of increasing the number of grid points $T$ for $y(t)$ from 30 to 60
is not shown, as it decreased relative errors for all quantities by a factor of
about 0.7 to 0.5.
Baseline rIMSE values (top left panel) are given for data with
$\snr_\epsilon=1$, $M=10$, $n_i=3$, $T=30$, $\snr_B=0.2$. 
In this
very noisy setting with small sample size and dominant random effects,
covariate effect estimates are not very accurate, with relative errors
mostly in the vicinity of one.
Since the random effects are estimated with little error, however, the error for
the responses in this difficult setting is small as well.
Increasing $\snr_\epsilon$ from 1 to 5 (top right
panel) decreases relative errors about 16-fold, with smaller 8-fold reductions
for the random effects in scenario 1.
Increasing  the number of groups from $M=10$ to $M=100$ 
(second row, left panel) has no substantial effect on the overall
estimation accuracy of  the $y_i(t)$.
Estimation accuracy of the functional random effects is not improved either due to the commensurate increase in the
number of parameters, while errors for the covariate
effects decrease about 8-fold. 
An increase in the average number of observations per group from $n_i=3$ to
$n_i=20$ (second row, right panel) results in a similar reduction of relative
errors for the covariate effects, and also a marked four- to sixfold
decrease in the errors for the response trajectories. 
A reduction of the relative contribution of the random effects to the linear
predictor, i.e. increasing $\snr_B$ from 0.2 to 1 [5] 
(bottom row, left [right] panel), improves the overall estimation accuracy of
$y(t)$ only slightly  if at all [factor 0.7 to 0.8]. This
overall improvement is due to the large reduction of errors for the covariate effects, which 
compensates for the observed deterioration of random effect estimates. While
the errors for the former decrease about 8-fold [16-fold],
the errors for the latter increase about 1.5- to twofold [five- to 16-fold].

\paragraph{Comparison to other approaches}
Appendix \ref{app:simstudy} summarizes additional results for comparisons between spline-based and FPC-based terms implemented for function-on-function effects and functional random effects in \code{pffr} as well as the wavelet-based approach for functional linear mixed models of \citet{Morris:Carroll:2006} implemented in \code{WFMM} \citep{WFMM}.

\paragraph{Coverage}
We also evaluate coverage of approximate point-wise empirical
Bayes confidence intervals (CIs) \citep[c.f.][eq.~(4.35)]{Wood2006} with
constraint correction \citep{Marra2012} for a nominal level of 95\%.
For each fitted model, we record the proportion of point-wise intervals
covering the true value of each quantity evaluated on a fine grid.
Note that the coverages of neighboring grid points are not independent,
but for the computationally intensive models we fit 
this is a feasible alternative to coverage estimates 
based on hundreds of replicates of each setting.
 CI coverage was consistently very close to the nominal level for $\hat y(t)$
($\text{Median}_{\text{10\% quantile}-\text{90\% quantile}}:
0.95_{(0.92-0.97)}$) and $\hat b_0(t)$ ($0.95_{(0.9-0.98)}$), while $\hat
\alpha(t)$ ($0.97_{(0.9-1)}$) showed some overcoverage as well as a few
replicates with coverage below 0.7 for small and noisy data.
Coverage for functional random slopes $\hat b_1(t)$ was below nominal for small
groups, but close to nominal for larger datasets ($n_i=3$: $0.9_{(0.81-0.96)}$;
$n_i=20$: $0.95_{(0.9-0.99)}$). Similarly for $\hat \beta_1(s,t)$ and $\hat
\beta_2(s,t)$, overall coverage was close to the nominal level
($0.95_{(0.85-0.99)}$), with systematic undercoverage in small and noisy data
sets with dominating random effects. Both $\hat\gamma_1(z_{1}, t)$
($0.99_{(0.94-1)}$) and $\hat\delta_2(t)$ ($1_{(0.75-1)}$) had overcoverage, the
latter with many outliers with observed coverages below $0.8$. 

\paragraph{Computation times}

Figure \ref{fig:sim:comptimes} shows computation times on an 2.2 GHz
AMD Opteron 6174 processor for the different scenarios and
sample sizes.  Especially for models with multiple random effects
(scenario 1) computation times increase dramatically in $M$.
Smaller models are fit rapidly, and even for the largest 
data sets with $nT=1.2\cdot10^5$, computation times are not 
prohibitively long. Speed gains for REML inference on large data sets
can be achieved by using the \code{pffr()}-option to use \pkg{mgcv}'s
\code{bam()} routine for estimating additive models on data sets that
do not fit into memory, as in Section \ref{sec:empeval:dti}.  
Using GCV optimization \citep{Wood2004} instead of REML-based inference
in \code{pffr()} can yield up to 10-fold speedups especially for large data sets,
but tends to be less stable.

\paragraph{Summary}
Important effects that contribute relevantly to the
predictor are estimated with good to excellent accuracy. 
Only a single replicate resulted in an rIMSE for
$y(t)$ greater than $0.1$ -- even in the most challenging data situations with few noisy
observations and small group sizes, our approach is able to reproduce the true
structure of the data well. 
Our results indicate that estimation accuracy of
covariate effects is affected most strongly by changes in the noise levels 
$\snr_\eps$ and especially
$\snr_B$, and less strongly by changes in the available number of observations
$M, n_i$ and $T$. 
The patterns of relative change in accuracy are identical for simple functional regression coefficients,  index-varying smooth
effects or effect surfaces for functional covariates. 
The estimation accuracy of the functional
random effects is affected  strongly by the relative importance of
the random effects $\snr_B$ and the group size $n_i$, and little by the  number of groups $M$.  
FPC-based random effects seem to require a sufficiently large number of groups and low noise level 
 to obtain usable FPC estimates.
Spline-based approaches yielded superior results to FPC-based and wavelet-based
approaches, but it should be noted that the data-generating process for the simulation study was spline-based itself.     
Overall, the observed coverage of the approximate
pointwise CI was very close to the nominal level except for very small or noisy
data.

\subsection{Modeling spatial association of
demyelination in a longitudinal MS study}\label{sec:empeval:dti}

\begin{figure}[htbp]
\begin{center}
\includegraphics[width=\textwidth]{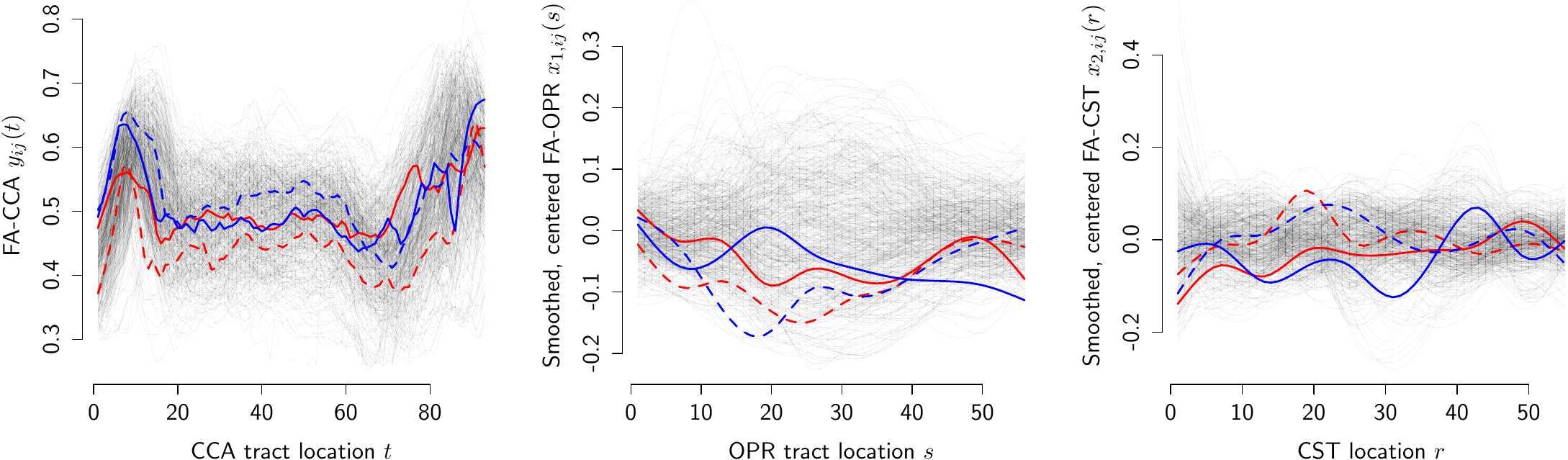}
\caption{From left to right: FA profiles along CCA, OPR and CST for MS
patients (red) and controls (blue). Solid line: females; dashed: males.
FA-OPR and FA-CST are de-trended and smoothed.}
\label{fig:data}
\end{center}
\end{figure}

Our motivating tractography study comprises 162 MS patients and 42 healthy controls who are observed
at one to eight visits, spread over up to four years, with 476 visits in total. 
MS damages white matter tracts (WMT)
in the brain due to lesions, axonal damage and demyelination. Diffusion tensor imaging
(DTI)  is a magnetic resonance imaging technique that is able to resolve
individual WMTs in the central nervous system \citep{Basser:2000}, and is thus a very useful tool in monitoring
disease progression in MS patients. At each visit, fractional anisotropy (FA) was determined via DTI along the
corpus callosum (CCA, connecting the left and right hemispheres of the brain), 
the left corticospinal tract (CST, connecting the brain and the spinal cord),
and the left optic radiation tract (OPR, connecting visual cortex and thalamus).
FA is derived from the estimated diffusion tensor and is equal to zero
if water diffuses perfectly isotropically (Brownian motion) and to one if it
diffuses  with perfectly organized  movement of all
 molecules in one direction for a given voxel. 
It may be decreased in MS patients and thus serves as a marker of
disease progression here.  Tracts are registered within and between subjects using standard biological
landmarks identified by an experienced neuroradiologist. 
Figure \ref{fig:data} displays
registered tract profiles as functions of tract location; 
profiles corresponding to four different subjects at first visit 
are highlighted.

Various aspects of this complex tractography dataset
have been explored in a sequence of papers including
\citet{Goldsmith:etal:2011}, \citet{Goldsmith:etal:2012},
\citet{Staicu:Crai:Ruppert}, \citet{Ivanescu2011}. 
This study was first introduced by \citet{Greven2010}, who  modeled
longitudinal variability in trajectories FPC-based, but could not take
into account any covariate effects.
Our goal here is to better understand the spatial course of the
demyelination process via its FA proxy and investigate possible differences
therein between MS and healthy subjects. 
\citet{Ivanescu2011}
considered a similar question, but used only the first measurement of each
subject since their approach is unable to handle the longitudinal structure of
the data. 
We assume a functional linear dependence between the FA along the CCA 
and the two functional covariates -- FA along the OPR and FA along the CTS
-- while adjusting for the effects of other relevant covariates such as gender,
age at visit, and disease status. Specifically, if $y_{ij}(t)$ is the FA
profile at location $t$ on the CCA tract observed at visit $j$ 
for subject $i$, we assume that the conditional mean of $y_{ij}(t)$, $\mu_{ij}(t)$, is  
\begin{equation}
\label{eqn:DTImodel0}
\begin{split}
\hspace{-.5cm}  \mu_{ij}(t)=  \alpha_{d_i}(t)  + \delta_{g_i}(t) + \nu(u_{ij}, t)+
            \int x_{1,ij}(s) \beta_{1, d_i}(s,t) ds +\int x_{2,ij}(r)
            \beta_{2, d_i}(r,t) dr 
\end{split}
\end{equation}
where $x_{1,ij}(s)$ and $x_{2, ij}(r)$ are the FA profiles at locations $s$ and $r$ along
the OPR and CTS tracts, respectively, observed at the $j$th visit of the $i$th subject.
Here $d_i$ is the disease status of the subject, with $d_i=1$ for MS patients, and $d_i=0$ for
healthy subjects; $g_i$ indicates the gender: $g_i=1$ for males and $0$ for females; and
$u_{ij}$ is the age (in years) at the $j$th visit of the $i$th subject. 
Note that the effects of FA-OPR and FA-CST at the current visit are
disease group-specific, with $\beta_{1,0}(s,t),
\beta_{2,0}(r,t)$ for controls and $\beta_{1,1}(s,t), \beta_{2,1}(r,t)$ for MS patients.
Neither age nor gender effects were found to differ between disease groups in the model-building process.

Effect estimates for a na\"ive model \eqref{eqn:DTImodel0} along the lines of \cite{Ivanescu2011} under assumed independence with measures of uncertainty are provided in Appendix \ref{app:dti}, Figure \ref{fig:dti-m-pffr-1-2} for completeness.  Due to the inappropriate conditional independence assumption, this
approach underestimates the variability of the estimates. 
We use our proposed functional additive mixed model to account for the 
within-subject correlation, which is the key advantage of our approach
over available function-on-function regression methods. Specifically, a more appropriate model is 
\begin{equation}
\label{eqn:DTImodel1}
\begin{split}
  y_{ij}(t) &=  \mu_{ij}(t) + b_{i0}(t) +\eps_{ijt},
\end{split}
\end{equation}
where $b_{i0}(t)$ are subject-specific functional random intercepts.
Model \eqref{eqn:DTImodel1} can be
fit using the \code{pffr()} function in the \pkg{refund} package. 
Estimating \eqref{eqn:DTImodel1} took about 17 hours on an 2 GHz AMD
Opteron processor. 

Since there are subjects 
with a few missing locations along the tracts  and since 
the FA measurements are observed with noise, we preprocess the
functional covariates. (Note that
missing values in the functional response are  not an issue for our
approach.)
The FA profiles are first detrended by subtracting the disease group-specific mean function
to make the estimated effects easily interpretable (see Appendix
\ref{sec:practicals:constr}). They are then smoothed, which also imputes missing values. For smoothing, we use functional principal component
analysis \citep{Di2009, Yao2005} for all tract-specific FA
curves under a working independence assumption between profiles 
on the same subject. Since the observed FA-CCA profiles exhibit a lot of small scale structure 
at locations $5-20$ and $>85$, spline based functional random intercepts would
require a very large basis   to provide sufficient flexibility.
Instead, we use the residual curves from model \eqref{eqn:DTImodel0} fitted
under an independence assumption to obtain an unsmoothed FPC-basis for the random intercepts,
as described on page~\pageref{sec:penreg:pcre}. 

\begin{figure}[htbp]
\includegraphics[width=\textwidth]{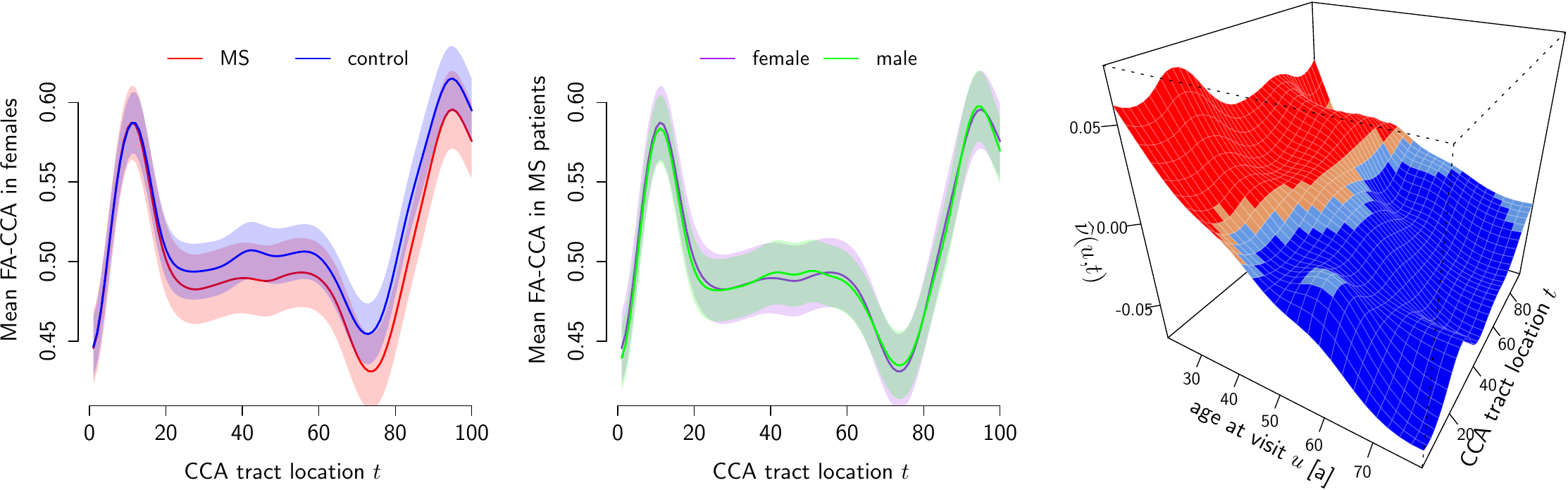}
\caption{Estimated components of model (\ref{eqn:DTImodel1}) with $\pm$2
pointwise standard errors. Coefficient surfaces are color-coded for sign and
approximate pointwise significance (95\%):
blue if sig. < 0, light blue if < 0, light red if > 0, red if sig. > 0.
Left to right: mean FA-CCA 
for healthy (blue, dotted) versus MS (red, solid) females; 
mean FA-CCA for female (purple, solid) and male 
(green, dotted) MS patients; estimated smooth index-varying age effect 
$\widehat\nu(u, t)$.}
\label{fig:dti-m-1}
\end{figure}
\begin{figure}[htbp]
\includegraphics[width=\textwidth]{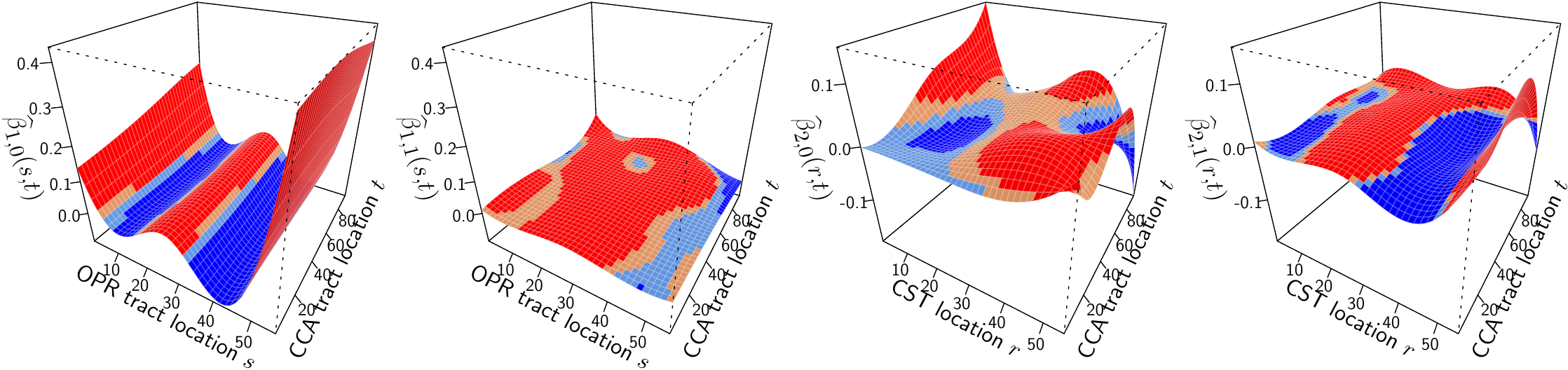}
\caption{Left to right: Estimated coefficient surfaces $\widehat\beta_{1,0}(s,
t)$, $\widehat\beta_{1,1}(s, t)$, $\widehat\beta_{2,0}(r, t)$,
$\widehat\beta_{2,1}(r, t)$.}
\label{fig:dti-m-2}
\end{figure}
\begin{figure}[htbp]
\centering
\includegraphics[width=.5\textwidth]{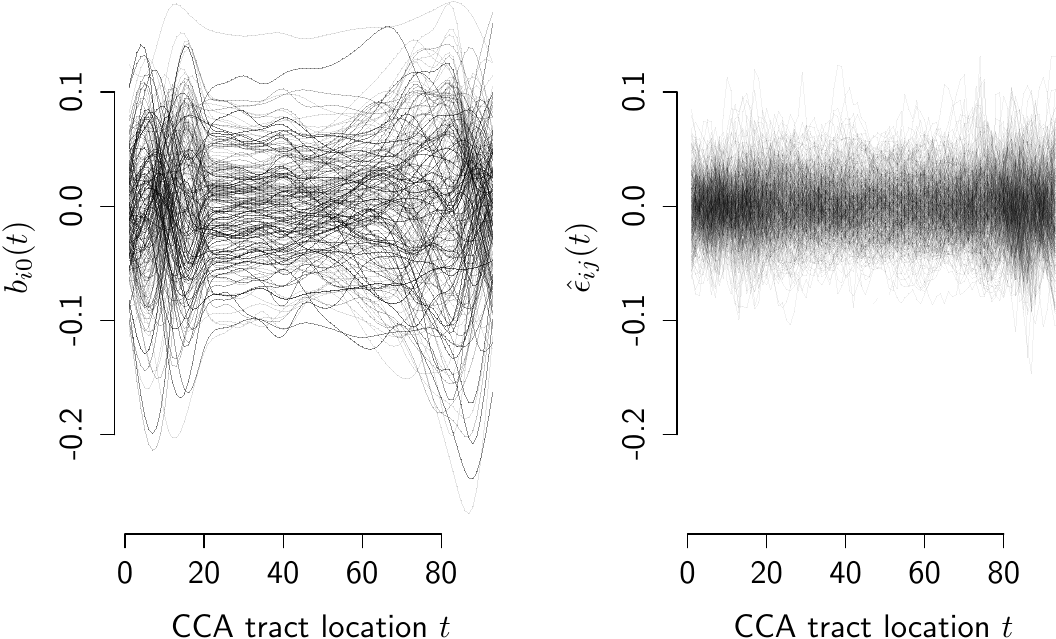}
\caption{Predicted functional intercepts $\hat b_{i0}(t)$ and observed
residuals $\hat \epsilon_{ij}(t)$ for
model (\ref{eqn:DTImodel1}).}
\label{fig:dti-m-3}
\end{figure}
Figure \ref{fig:dti-m-1} shows the estimated mean of the FA profiles along the CCA tract (anterior to posterior, i.e.~front  of the head to the back)
for female subjects with and without MS (left
panel) and for male and female MS patients (second from left).
The estimated mean FA profiles have similar shapes, with a sharp increase
in the rostrum/genu (front), a plateau in the middle section, followed by a decrease near the isthmus and a rapid
increase towards the splenium (back). 
As expected, MS patients tend to have lower FA-CCA, especially in the posterior section from the rostral body to the splenium. 
The effect of gender seems to be negligible.
The estimated age effect, $\hat \nu(u_{ij}, t)$, indicates that FA-CCA decreases 
almost linearly with age over the entire tract, particularly in the anterior part, but this effect is fairly small.
Not accounting for the longitudinal data structure 
(c.f. Figure
\ref{fig:dti-m-pffr-1-2}), differences between MS and healthy subjects would be found to be much larger and statistically significant along the entire tract. 
The corresponding estimate for the age effect  seems implausible. Due to the misplaced independence
assumption, the variability of the estimates shown in Figure
\ref{fig:dti-m-pffr-1-2} is underestimated, but should
be approximately correct in Figures \ref{fig:dti-m-1} and \ref{fig:dti-m-2}.
The rightmost panels in Figure \ref{fig:dti-m-pffr-3-4} give covariances and correlations for $\eps_{ij}(t)$ and
show that the white-noise-error assumption is reasonable for model \eqref{eqn:DTImodel1}, 
but severely violated for \eqref{eqn:DTImodel0}. They also show that
spline-based random intercepts are less successful in removing all structure
from the residuals in this case, especially in the rostrum/genu.

In healthy controls, FA values at the ends of the OPR tract (towards lateral geniculate nucleus and visual cortex, respectively) and in its middle section show a positive association with FA values along the entire CCA tract (see Figure \ref{fig:dti-m-2}). 
For the CTS tract, there is some indication of a positive association between FA values in the beginning of the CTS tract (medulla) and the end of the CCA (splenium) and between the end of the CTS tract (subcortical white matter) and the beginning of the CCA (rostrum/genu), the latter corresponding with spatial proximity. 
These patterns should be indicative of the normal ageing process, while the observed associations mostly vanish for MS patients or become much weaker. It should be noted, however, that simulation
results  indicate  potentially low  estimation accuracy of fixed effects  in
 settings such as this one in which the random effects dominate the predictor.
Figure \ref{fig:dti-m-3} displays the predicted intercept curves $\hat
b_{i0}(t)$ (left panel) and observed residuals $\hat \epsilon_{ij}(t) = y_{ij}(t)-\hat\mu_{ij}(t)-\hat
b_{i0}(t)$ (right panel).
The large variation in the predicted functional intercepts reveals large inter-subject variability. 
By accounting for the between-subject variability the observed integrated root
mean square error of the responses with the proposed method (0.027) reduces to half
of its magnitude compared to  \citet{Ivanescu2011} under an independence assumption (0.05).
Model \eqref{eqn:DTImodel1} explains about $90\%$ of the observed variability,
while \eqref{eqn:DTImodel0} explains only about $63\%$.

In conclusion, using our flexible
modeling framework for the FA profiles along the CCA tract  shows that a large fraction of the variability in the data is captured by
subject-specific random effects. Modeling the dependence on FA profiles at other well
identified tracts, OPR and CTS, can provide new insights into the spatial association in normal ageing and disease processes. Interestingly, our results indicate that the
associations between 
demyelination along the left CTS and left OPR tracts and the CCA tract are weaker for MS patients
than for healthy controls. A possible interpretation of
this finding could be that demyelination processes in MS patients are more
strongly localized, consistent with the development of localized lesions during MS. 
By properly accounting for the longitudinal
structure of the data the estimation uncertainty of all  effects increases compared to model \eqref{eqn:DTImodel0} under an independence assumption. 


\section{Discussion and Outlook}\label{sec:finis}
We propose a general framework for flexible additive regression models for correlated functional responses, 
allowing for multiple functional random 
effects with, for example, spatial, temporal, spatio-temporal or longitudinal (Section
\ref{sec:empeval:dti}) correlation structures. Dependence structures can be modeled either
implicitly by specifying smooth temporal, spatial or tempo-spatial effects
or explicitly by including functional random effects with marginal between-unit correlation structures
given by the precision matrices of Gaussian (Markov) random fields. Estimation and inference is performed by
standard additive mixed model software, allowing us to take advantage of established robust and
flexible  algorithms. The approach is
implemented as fully documented open-source software in the
\code{pffr()}-function in the \pkg{refund} package \citep{refund} for \R.
Effects of functional covariates and functional random effects are available in
both FPC- and spline-based variants and both scalar and functional covariates
can have linear or more general smooth effects on the outcome trajectories,
allowing analysts to choose the most suitable tools for the task at hand.

Simulation experiments show that the proposed method recovers relevant
effects reliably and handles small group sizes and/or low numbers of
replications well. Data sets of considerable size can be fit in
acceptable time. Two applications demonstrate that our approach makes it possible to fit flexible models that do justice to
complex data situations and yet still yield interpretable results that can help 
to understand the underlying processes.  
 
This work opens up a number of interesting avenues for further research.
A first direction concerns the covariance structure
of the residuals. Since our present  inference algorithms do not
exploit the extreme sparsity of the design matrices for smooth observation-specific residual terms, 
estimating such terms dramatically increases 
computation time and memory requirements for large data
sets. On the other hand, simply assuming $\iid$
errors $\epsilon_{it}$ will often be unrealistic since 
some degree of auto-correlation and heteroscedasticity over the index of
the functional response is usually encountered in practice. 
We are currently investigating an iterative procedure similar to
the approach in \citet{Reiss2010}, where observed residuals from an
initial model estimated under a working independence assumption are
used to estimate a working covariance structure and the model
 is then re-estimated based on de-correlated data. If successful,
such a marginal model specification could offer an
efficient alternative to the conditional modeling approach outlined in the
present paper. In a second direction, we are currently developing diagnostic measures 
to identify potential problems with low-rank functional covariates
(c.f.~Appendix \ref{sec:practicals}) as well as practical model-building strategies 
regarding the estimation of corresponding regression surfaces.
A closely related avenue of inquiry are more in-depth comparisons of spline-
and FPC-based approaches for modeling function-on-function terms as well as functional random effects
in order to evaluate their relative strengths and weaknesses.
The unifying framework implemented in \code{pffr()} will greatly facilitate
such comparisons. In addition, we have begun implementing a dedicated toolbox for
REML-based inference tailored to function-on-function regression. 
This effort is based on the computationally efficient array regression approach
of \citet{Currie2006}, which is expected to speed up inference for large scale problems
and help to generalize the proposed methods for multidimensional  
functional responses and image regression.

\section*{Acknowledgements}
We thank Danny Reich and Peter Calabresi for supplying and explaining the DTI
tractography data. Richard Herrick and Jeffrey S. Morris provided a Linux-version of \code{WFMM}. Sonja Greven and Fabian Scheipl were funded by Emmy Noether
grant GR 3793/1-1 from the German Research Foundation. Ana-Maria Staicu's research was supported
by U.S. National Science Foundation grant number DMS 1007466.  

\begin{scriptsize}
\setlength{\bibsep}{0em}
\bibliography{lfpr-arxiv}
\end{scriptsize}

\begin{small}
\appendix
\section{Identifiability}\label{sec:practicals}

\subsection{Imposing suitable identifiability
constraints}\label{sec:practicals:constr} 
Additive models for scalar responses
 ensure identifiability by imposing suitable constraints on the functions that make 
up the additive predictor \mbox{$\beta_0 + \sum_s f_s(x_s)$}, such as  a sum-to-zero constraint $\sum^n_{i=1}
f_s(x_{si})=0$ for each function $f_s(x_s)$  \citep{Wood2006}. Otherwise, any constant could be added to one function and subtracted from the others without changing the fit criterion. 

A similar issue arises in the context of our proposed model.
For arbitrary functions $\bar{\gamma}_{z}(t)$, $\bar b_{g}(t)$
\begin{align*} 
\EV(y_{i}(t)) &= \alpha(t) + \gamma(z_{i}, t) + b_{g(i)}(t) = (\alpha(t)  + \bar{\gamma}_{z}(t) + 
\bar b_{g}(t)) + (\gamma(z_{i}, t) - \bar{\gamma}_{z}(t)) + (b_{g(i)}(t) - \bar
b_{g}(t))
\end{align*}
obtains the same fit with two different parameterizations. To avoid this,
we impose sum-to-zero constraints for each $t$ so that  $ n^{-1}
\sum^n_{i=1}  b_{g(i)}(t) = n^{-1} \sum^n_{i=1} \gamma(z_{i}, t) = 0$ $\forall t$.

We also center covariate trajectories
 $x_{i}(s)$ by subtracting the mean function $\bar x(s) = n^{-1} \sum_{i}  x_{i}(s)$.
If both the sum-to-zero constraints for each $t$ are imposed and functional
covariates are centered, all effects that
vary over the index of the response are directly interpretable as deviations from the overall
mean trajectory $\alpha(t)$.
Standard sum-to-zero constraints implemented in
\pkg{mgcv}, which would correspond to $\sum_{i,t} \gamma(z_i, t) =
0$, yield neither identifiable models nor effects that are
interpretable like this. Implementationwise, we use the method described in \citet[ch. 1.8.1]{Wood2006}
to absorb the sum-to-zero-for-each-$t$ constraints into the design matrices of
all effects varying over $t$, see section A in the online supplement for details
and examples.

\subsection{Limits on the identifiability of complex regression surfaces for
low-rank functional covariates}\label{sec:practicals:ident} 
For function-on-function-regression terms $\int_\mathcal{S}
x(s)\beta(s,t)ds$,
identifiability of $\beta(s,t)$ is guaranteed under 
conditions derived in \citet{HeMullerWang2003}, 
\citet{ChiouMullerWang2004} and \citet{PrchalSarda2007}, 
which are hard to verify empirically. 
In practice, an important quantity in this regard for the stability of spline-based estimates 
is the effective rank of the covariance operator of $x(s)$, which can be defined as the number of eigenvalues
that together account for at least 99.5\% of the covariate's variability.
If this effective rank is low, the kernel of the functional
covariate's covariance operator is large. 
\citet{ScheiplGreven2012} have shown that
spline-based regression surface estimates can be unstable if the kernel of the functional
covariate's covariance operator overlaps the function space spanned by parameter
vectors in the nullspace of the tensor product spline's roughness penalty. 
Based on theoretical considerations and simulation results
\citep[c.f.~][]{ScheiplGreven2012}, we recommend that practitioners check the
effective rank of the observed covariance matrix of functional covariates
and the amount of overlap between the kernel of the functional
covariate's covariance operator and the nullspace of the associated roughness
penalty. Utility functions to perform these checks and constructors for modified roughness
penalties without nullspaces are included in \pkg{refund}.

\section{Supplementary Details for the DTI Data Analysis}\label{app:dti}

\begin{figure}[htbp]
\centering
\includegraphics[width=.8\textwidth]{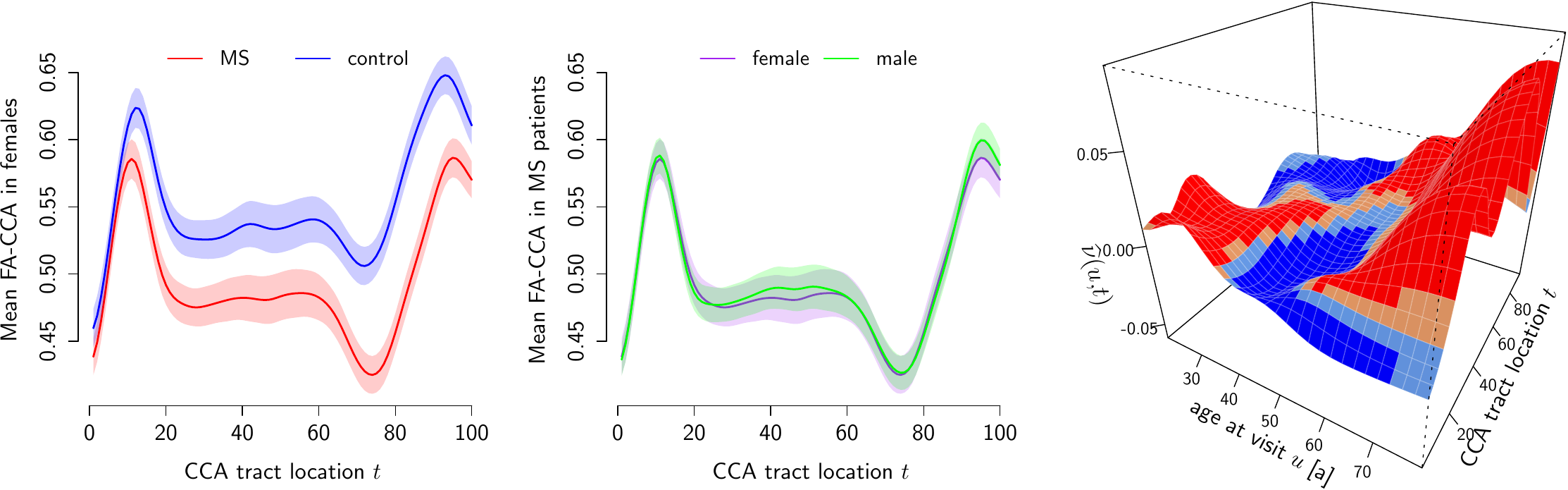}
\vskip 1em
\includegraphics[width=.8\textwidth]{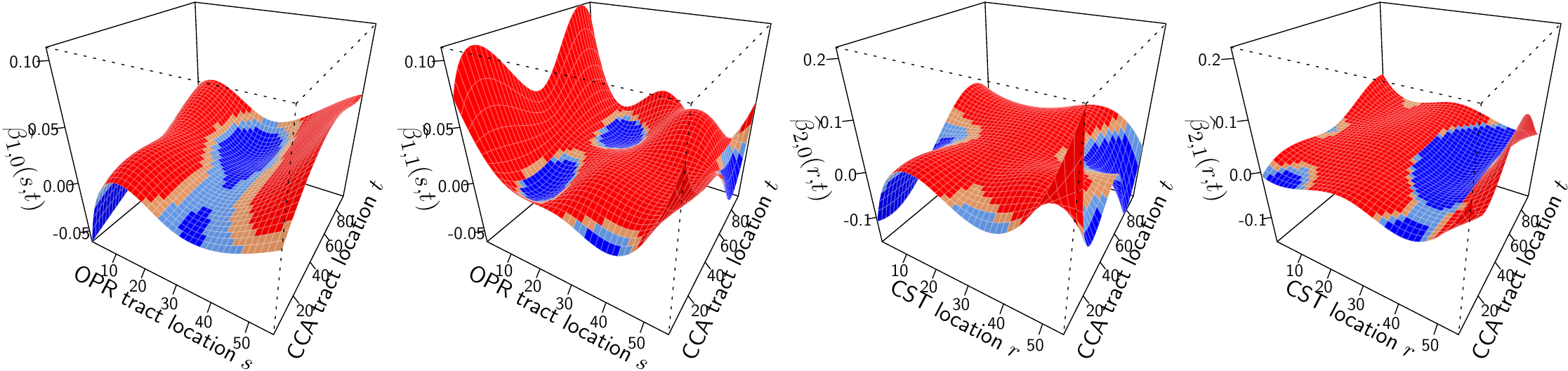}
\caption{Estimated components of model \eqref{eqn:DTImodel0} with $\pm$2
pointwise standard errors, using \cite{Ivanescu2011}. Coefficient
surfaces are color-coded for sign and pointwise significance (95\%):
blue if sig. $< 0$, light blue if $< 0$, light red if $> 0$, red if sig. $> 0$.
Top row, left to right: mean FA-CCA 
for healthy (blue, dotted) versus MS (red, solid) females; 
mean FA-CCA for female (purple, solid) and male 
(green, dotted) MS patients; estimated effect of age-at-visit  $\widehat\nu(u, t)$.
Bottom row, left to right: Estimated coefficient surfaces $\widehat\beta_{1,0}(s,
t)$, $\widehat\beta_{1,1}(s, t)$, $\widehat\beta_{2,0}(r, t)$,
$\widehat\beta_{2,1}(r, t)$.}
\label{fig:dti-m-pffr-1-2}
\end{figure}
\begin{figure}[htbp]
\centering
\includegraphics[width=.8\textwidth]{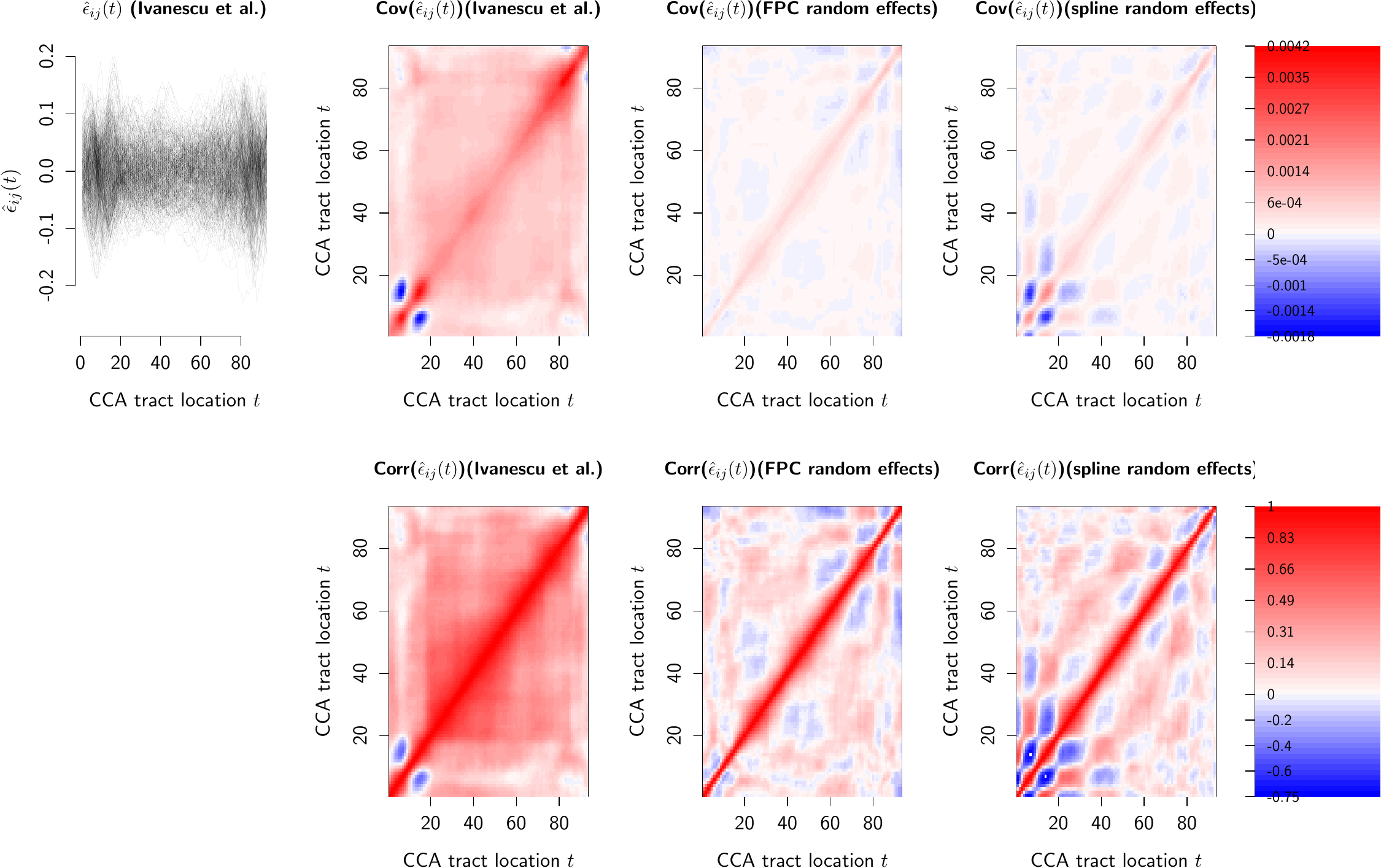}
\caption{Top row, left to right: Observed residuals $\hat \epsilon_{ij}(t)$ for
model (\ref{eqn:DTImodel0}); empirical covariance for $\hat \epsilon_{ij}(t)$ for
model (\ref{eqn:DTImodel0}); empirical covariance for $\hat \epsilon_{ij}(t)$ for
model (\ref{eqn:DTImodel1}) with FPC-based random intercepts; empirical covariance for $\hat \epsilon_{ij}(t)$ for
model (\ref{eqn:DTImodel1}) with spline-based random intercepts; legend for
covariance values. Bottom row: Empirical correlations.}
\label{fig:dti-m-pffr-3-4}
\end{figure}

\clearpage
\section{Supplementary Simulation Study Results}\label{app:simstudy}

\paragraph{Comparison with FPC-based approaches}
We fit models with an FPC-based function-on-function term (c.f.
page \pageref{sec:penreg:ffpc}) and models with FPC-based functional random
intercepts (c.f.
page \pageref{sec:penreg:pcre}) to each dataset generated for the second scenario. 
rIMSEs for the FPC-based function-on-function term were larger than those of the spline-based estimates
by a mean factor of $1.5_{(1.2-2.1)}$, while
computation time was about the same for $M=10$ ($1_{(0.8-1.4)}$) and
somewhat longer for $M=100$ ($1.3_{(1.1-1.6)}$).  
Results for the FPC-based functional random intercept were more different from
the spline-based option. Specifically, the FPC-based functional random intercept
showed fairly little improvement for $\snr_\epsilon=5$ compared to 
$\snr_\epsilon=1$. For the latter, the FPC performance was fairly similar
($M=10$: factor of $2.4_{(1.3-9.4)}$, $M=100$: factor of
$1.1_{(1.0-1.5)}$), while it was much less precise for the former:  $3.4_{(1.1-17)}$ for $M=100$ and
$17_{(2.5-106)}$ for $M=10$. As expected, however, FPC-based functional
random intercepts scaled much better than spline-based ones for larger datasets
in terms of computation time due to their more compact optimal basis
representation -- for $M=100$, the  iterative FPC procedure was faster than 
spline-based random effect models by
a factor of $0.3_{(0.2-0.5)}$. Also,  our spline-based 
data generating process corresponding to five non-zero FPCs 
(c.f.~Appendix B of the supplement) may be more difficult for FPC based
approaches: previous simulation studies of FPC-based functional regression have
typically used data generating processes with lower effective rank
\citep[e.g.][with 2, 3, and 4 eigenfunctions, respectively]{Muller2008, Wu2010,
Chen:Mueller:2012} and simpler coefficient shapes. 

\paragraph{Comparison with \code{WFMM}}
We compare our approach to the available implementation of the wavelet-based functional linear mixed
models of \citet{Morris:Carroll:2006} in \code{WFMM} \citep{WFMM}. We can only provide this comparison
for scenario 1 as the other scenarios feature terms that are not available in \code{WFMM}, which can
only fit random effect curves and functional linear effects $z_{ij} \beta(t)$ of scalar covariates $z$. 
Note that, differing from the results for \code{pffr} in the remainder of the article, these results are
for balanced data, as the \code{WFMM} algorithm seems to fail whenever there are any subjects with 1 or
2 observations only, and 10 replicates per setting. We used the default hyper- and tuning parameters for
\code{WFMM}, with 2000 iterations of burn-in followed by 10000 iterations of sampling.  In general, the
IMSEs for \code{WFMM} are about double to three times those of \code{pffr}. Specifically, the IMSE for
$y(t)$ is increased by a median factor of $2.5_{(2.1-3.1)}$ (IMSE($b_0(t)$: $2.3_{(1.6-3.8)}$,
IMSE($b_1(t)$: $2.4_{(1.5-4)}$).  Note, however, that this comparison is not entirely fair to
\code{WFMM}, as it is designed for spiky data e.g. from spectrometry (i.e., it assumes sparsity in a suitable
wavelet domain), not the smooth functional data that we assume and correspondingly simulated here. Although \code{WFMM} is much
slower (4- to 16-fold) than \code{pffr} for small and intermediate data sizes, its computation time
increases much slower than \code{pffr} for larger data sets due to its efficient data representation in
the wavelet domain and its very fast \code{C++} implementation. Detailed results for this comparison are
provided in Section B.6 of the online appendix.

\section{Online Supplement}
Supplementary material with extensive code examples is available from the first author's homepage at \url{http://www.statistik.lmu.de/~scheipl/research.html}.

\end{small}
\clearpage

\end{document}